\documentclass[doublecol]{epl2} 

\newcommand{\be}{\begin{equation}}
\newcommand{\ee}{\end{equation}}

\newcommand{\bea}{\begin{eqnarray}}
\newcommand{\eea}{\end{eqnarray}}
\newcommand{\bd}{\begin{displaymath}}
\newcommand{\ed}{\end{displaymath}}
\newcommand{\ba}{\begin{array}}
\newcommand{\ea}{\end{array}}
\newcommand{\bi}{\begin{itemize}}
\newcommand{\ei}{\end{itemize}}
\newcommand{\bc}{\begin{center}}
\newcommand{\ec}{\end{center}}
\newcommand{\bfl}{\begin{flushleft}}
\newcommand{\efl}{\end{flushleft}}
\newcommand{\bfr}{\begin{flushright}}
\newcommand{\efr}{\end{flushright}}

\def\ket#1{\left\vert #1 \right\rangle}

\def\6{\partial}

\def\bra{\langle}
\def\ket{\rangle}
\def\={\!\!\!&=&\!\!\!}
\def\+{\!\!\!&&\!\!\!+~}
\def\-{\!\!\!&&\!\!\!-~}

\newcommand*{\auf}{\uparrow}
\renewcommand*{\vec}[1]{\mathbf{#1}}

\title{Theory of nonequilibrium dynamics  of multiband   superconductors}

\author{Alireza Akbari\inst{1,2} \and Andreas P. Schnyder\inst{3} \and  Dirk Manske\inst{3} \and Ilya Eremin\inst{2}}
\shortauthor{A. Akbari \etal}

\institute{                    
  \inst{1} Max-Planck Institute for the  Chemical Physics of Solids, D-01187 Dresden, Germany \\
    \inst{2}  Theoretische Physik III, Ruhr-Universit\"{a}t Bochum, D-44780, Bochum, Germany\\
   \inst{3}  Max-Planck-Institut f\"ur Festk\"orperforschung, D-70569 Stuttgart, Germany\\ 
  }
\pacs{74.40.Gh}{Nonequilibrium processes in superconductivity}
\pacs{78.47.J-}{Femtosecond techniques; in spectroscopy of solid state dynamics}
\pacs{78.20.Bh}{Optical properties; theory and models}

\abstract{
 We study the nonequilibrium dynamics of multiband BCS superconductors subjected to ultrashort pump pulses.
Using density-matrix theory,
the time evolution of the Bogoliubov quasiparticle densities and the superconducting order parameters are computed
as a function of pump pulse frequency, duration, and intensity.
Focusing on two-band superconductors, we consider two
different model systems. The first one,
relevant for iron-based superconductors, describes
two-band superconductors with a repulsive interband interaction $V_{12}$ which is much larger
than the intraband pairing terms.
The second model, relevant for MgB$_2$, deals with the opposite limit where
the intraband interactions are dominant and the interband pair scattering
$V_{12}$ is weak but attractive.
For ultrashort pump pulses,  both of these models
exhibit a nonadiabatic behavior which is characterized by oscillations of the superconducting order parameters.
We find that for nonvanishing $V_{12}$, the superconducting gap on each band exhibits two oscillatory frequencies which are
determined by the long-time asymptotic values of the  gaps.
The relative strength of these two frequency components depends sensitively on the magnitude of the
interband interaction $V_{12}$.
}

\begin{document}

\maketitle

The discovery of  iron-based superconductors\cite{kamihara} and MgB$_2$\cite{akimitsu},
where superconductivity is characterized by
more than one order parameter, has lead to a
renewed interest in multiband superconductors.
The existence of multigap superconductivity was predicted more than fifty years ago
in the pioneering works by Moskalenko\cite{moskalenko} and Suhl \textit{et al.}\cite{suhl},
where it was argued that a band-dependent interaction potential can give rise
to different order parameters on different Fermi surface sheets.
The presence of multiple superconducting order parameters
has many interesting physical consequences, most of which are related  to the relative
amplitudes and phase differences between the gaps on different Fermi surfaces.

In recent years, numerous studies of the
nonequilibrium dynamics of multiband superconductors
have been performed using femtosecond time-resolved
spectroscopy\cite{demsar,demsar2,mansart09,mansart,uwe,torchinsky}.
The relaxation kinetics measured in these experiments gives
important information on the electronic band structure,
on the electron-phonon coupling strengths,
as well as on the symmetry of the superconducting order parameters.
For example, time-resolved measurements on the iron pnictide superconductor Ba$_{0.6}$K$_{0.4}$Fe$_2$As$_2$\cite{torchinsky} have revealed two distinct quasiparticle relaxation processes: a fast one, whose decay rate increases
linearly with excitation density, and a slow one, which is independent of  excitation density.
This behavior has been attributed to the multigap structure of the superconductor.
Furthermore, from a careful analysis of the temperature dependence of the decay rates, the authors
of Ref.~\cite{torchinsky} have been able to deduce the pairing symmetries of the superconducting state.

 In this paper, we shall not be concerned with the relaxation dynamics,
but rather with the nonequilibrium evolution of multiband superconductors at times shorter than the  relaxation time, i.e., with the nonadiabatic behavior of these system.
It has been shown that the nonequilibrium response at these short time scales exhibits interesting properties, such as order parameter oscillations\cite{volkov,yuz05,Yuzbashyan06,papenkort}.
In a pioneering work, B.~Mansart \textit{et al.}\cite{mansart11} have been able to detect these
 coherent oscillations of the order parameter in a high-T$_c$ cuprate superconductor via transient broad-band reflectivity measurements.
The recently discovered multiband superconductors MgB$_2$ and iron pnictide superconductors, which all have a relatively large
superconducting gap, offer an excellent opportunity to analyze this nonadiabatic dynamics also in multiband systems, 
where the interplay of the intraband pairing interactions and interband pair scattering terms play an important role. 
One goal of the present work, is to provide guidance for experimental studies of these interesting nonadiabatic response properties
in multigap superconductors. 

In the folowing, we employ the density-matrix theory
to study the temporal evolution of  Bogoliubov quasiparticle distributions
in multiband superconductors
after excitation by a short pump pulse.
We focus on the nonadiabtic regime, i.e., the regime where
the pump pulse duration $\tau$ is much shorter than the dynamical time scales of the superconducting order parameters.
In this case, the pump pulse creates rapid oscillations in the quasiparticle densities.
The essential physics of  multiband  superconductors can be
captured by a minimal two-band  model
with Hamiltonian $H=H_0 + H_{\scriptsize\textrm{int}}$, with
$H_0= \sum_{{\bf k}, l, \sigma }\varepsilon _{{\bf k} l}a_{{\bf k}l\sigma }^{\dagger }a^{\phantom{\dag}}_{{\bf k}l\sigma } $ and
\begin{eqnarray}
H_{\scriptsize\textrm{int}}
&=&
\sum_{{\bf k}, {\bf k}^{\prime },l}V^{ll}_{{\bf kk}^{\prime }}a_{{\bf k}^{\prime }l\uparrow
}^{\dagger }a_{-{\bf k}^{\prime }l\downarrow }^{\dagger }a^{\phantom{\dag}}_{-{\bf k}l\uparrow
}a^{\phantom{\dag}}_{{\bf k}l\downarrow}
\\
&& +\sum_{{\bf k}, {\bf k}^{\prime }, l\neq l^{\prime}} \left( V^{ll^{\prime}}_{{\bf kk}^{\prime }}a_{{\bf k}^{\prime
}l\uparrow }^{\dagger }a_{-{\bf k}^{\prime }l\downarrow }^{\dagger
}a^{\phantom{\dag}}_{-{\bf k}l^{\prime}\uparrow }a^{\phantom{\dag}}_{{\bf k}l^{\prime}\downarrow } + \textrm{h.c.}\right) ,
\nonumber
 \label{ham1}
\end{eqnarray}
where  $a_{{\bf k}l\sigma }$  is the  electron (hole)
annihilation operator with band index $l=1,2$, spin $\sigma$, and momentum~${\bf k}$.
The bare dispersions of the two bands, $\varepsilon _{{\bf k}1}$ and $\varepsilon_{{\bf k}2}$,
are assumed to be quadratic with $\varepsilon _{ {\bf k} l} = \hbar^2 {\bf k}^2 / ( 2 m_l) - E_{\scriptsize\textrm{F} l}$,
where $m_l$ and $E_{\scriptsize\textrm{F}  l}$ represent the effective electron (hole) mass and Fermi energy of band~$l$, respectively.
The intraband pairing interactions are denoted by
 $V^{11}_{{\bf kk}^{\prime }}$ and $V^{22}_{{\bf kk}^{\prime }}$, while the interband pair scattering is represented
 by $V^{12}_{{\bf kk}^{\prime }} = ( V^{21}_{{\bf k}^{\prime }  {\bf k}} )^{\ast}$.
Assuming that the splitting of the bands is much larger than the superconducting energy scales, we neglect
all interband pairing terms, i.e., we set  $\langle a_{-{\bf k}l\downarrow }a_{{\bf k} l^{\prime}\uparrow } \rangle =0$ for
all $l \ne l^{\prime}$.  Furthermore, we
restrict ourselves to momentum-independent pairing interactions with
$V^{11}_{{\bf kk}^{\prime }}=V_{11}$, $V^{22}_{{\bf kk}^{\prime }}=V_{22}$,
and $V^{12}_{{\bf kk}^{\prime }}=V_{12}=|V_{12}|e^{i\theta}$.
The intraband pair couplings are assumed to be attractive, i.e., $V_{11},V_{22} > 0$.
The phase $\theta$ of $V_{12}$ is taken to be either $0$ or $\pi$,
depending on whether the interband pair scattering is attractive or repulsive.

The dynamics of superconductors at times shorter than the
quasiparticle energy relaxation time can be fully described within mean-field BCS theory\cite{yuz05,volkov}.
We therefore perform a mean-field decoupling of the quartic interactions in Eq.~(\ref{ham1}),
which yields the BCS Hamiltonian
\begin{eqnarray} \label{MF_ham}
H_{\scriptsize\textrm{BCS}}
=
H_0 +
\sum_{ {\bf k} \in \mathcal{W} , l} \left(
\Delta_l a^{\dag}_{ {\bf k} l \uparrow}  a^{\dag}_{ -{\bf k} l \downarrow}
+
\Delta^{\ast}_l a^{\phantom{dag}}_{ - {\bf k} l \downarrow}  a^{\phantom{\dag}}_{ {\bf k} l \uparrow} \right), \quad
\end{eqnarray}
with the mean-field gaps
\begin{eqnarray} \label{MFgaps}
\Delta_{1} = \sum_{{\bf k} \in \mathcal{W} }
\left(
V_{11}\langle a_{-{\bf k}1\downarrow
}a_{{\bf k}1\uparrow }\rangle +V_{12}\langle a_{-{\bf k}2\downarrow }a_{{\bf k}2\uparrow
}\rangle
\right), \nonumber \\
 \Delta _{2} = \sum_{{\bf k} \in \mathcal{W} }
 \left(
V_{22}\langle a_{-{\bf k}2\downarrow }a_{{\bf k}2\uparrow }\rangle + V_{21}\langle
a_{-{\bf k}1\downarrow }a_{{\bf k}1\uparrow
}\rangle
\right) .  \label{gaph}
\end{eqnarray}
The sums in Eqs.~(\ref{MF_ham}) and (\ref{MFgaps}) are over the set $\mathcal{W}$ of
momentum vectors with $\left| \varepsilon_{\vec{k} l} \right|  \leq \hbar \omega_{\scriptsize\textrm{c}}$ and
$\omega_{\scriptsize\textrm{c}}$ the cut-off frequency.
The quasiparticle spectrum of $H_{\scriptsize\textrm{BCS}}$ is given by
$\left\{ - E_{ {\bf k} 1},  - E_{ {\bf k} 2}, + E_{ {\bf k} 1}, + E_{ {\bf k} 2} \right\}$,
where $ E_{{\bf k}l}=\sqrt{\varepsilon_{{\bf k} l}^{2}+|\Delta
_{l}|^{2}}$, with $l=1,2$.
In the equilibrium, the order parameters $\Delta_1$ and $\Delta_2$ are determined by
the self-consistency equation
\begin{equation} \label{gapeq}
\left(
\begin{array}{cc}
V_{11}K_1 & V_{12}K_2 \\
V_{21}K_1 & V_{22}K_2
\end{array}
\right)
\left(\begin{array}{c}
\Delta_1 \\
\Delta_2
\end{array}
\right)=
\left(\begin{array}{c}
\Delta_1 \\
\Delta_2
\end{array}
\right),
\end{equation}
where $ K_{l} = \sum_{{ \bf k} \in \mathcal{W}}
\tanh\left(E_{ {\bf k} l }/2T\right)/E_{{\bf k} l}$, with $l=1,2$.
The gap equation~(\ref{gapeq}) has a nontrivial solution whenever
the determinant of the corresponding secular matrix vanishes, i.e.,
 \begin{equation}
\det\left(%
\begin{array}{cc}
V_{11}K_1-1 & V_{12}K_2 \\
V_{21}K_1 & V_{22}K_2-1
\end{array}
\right)=0.%
\label{Keq}
\end{equation}
From this condition, the transition temperature $T_c$ can be obtained.

The two-band superconductor (\ref{MF_ham}) is perturbed by a short Gaussian pump pulse,
which in the Coulomb gauge is described by a transverse vector potential
$\vec{A}_{ {\bf q} } (t)=\vec{A}_0 e^{-\left( \frac{2 t\sqrt{\ln 2}}{\tau} \right)^2}
(\delta_{ {\bf q} , {\bf q}_0}e^{-i\omega_0 t}+\delta_{ {\bf q},- {\bf q}_0}e^{+ i\omega_0 t})$ ,
%
with amplitude $\vec{A}_0=A_0 \hat{\vec{e}}_y  $, photon wave vector $\vec{q}_0=q_0 \hat{\vec{e}}_x $, and full
width at half maximum $\tau$.
The optical pump pulse is centered at $t=0$ and has a
central frequency $\omega_0$.
The coupling of the vector potential ${\bf A}_{\bf q}$ to the two-band
superconductor is given by
\begin{eqnarray}
&& H_{\scriptsize\textrm{em}} (t)
=
         e\hbar
           \sum_{\vec{k},\vec{q},l,\sigma}
           \frac{ \vec{k}\cdot\vec{A}_{\vec{q}} (t)  }{m_l}a^{\dagger}_{\vec{k}
           + \vec{q} l \sigma}a^{\phantom{\dag}}_{\vec{k} l \sigma}
\\
&&\; \;  +
            e^2\sum_{\vec{k}, \vec{q} , l ,\sigma}
            \frac{1}{2 m_l}
    \left[        \sum_{\vec{q}'}
     \vec{A}_{\vec{q}-\vec{q}'} (t)  \cdot\vec{A}_{\vec{q}'}(t) \right]
           a^{\dagger}_{\vec{k}+\vec{q} l \sigma}a^{\phantom{\dag}}_{\vec{k}l \sigma} .
           \nonumber
\end{eqnarray}

\begin{figure}
  \centering
  \includegraphics[width=0.49\textwidth]{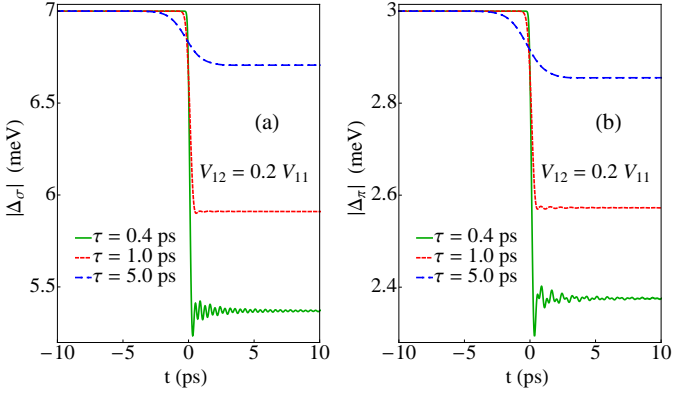}
\caption{(Color online)
Calculated time evolution of  $|\Delta_{\sigma}(t) |$ and $|\Delta_{\pi} (t) |$ for the case of MgB$_2$, for $V_{12}=0.2V_{11}$ and three different pump pulse widths $\tau=0.4$~ps (solid green),
$1.0$~ps (dotted red), and $5.0$~ps (dashed blue). The central energy of the pump pulse is $\hbar \omega_0=8.0$ meV, i.e., it is
slightly larger than $2 \Delta^{\pi} (t_i) $, but smaller than $2 \Delta^{\sigma} (t_i)$.
We take $A_0^2 \tau \approx 5.3 \times 10^{-27}
\frac{\textrm{J$^2$s$^3$}}{\textrm{C$^2$m$^2$}}$.
}
\label{fig1}
\end{figure}

By means of the density-matrix formalism, we numerically compute the coherent response of the above model Hamiltonian after excitation by a short pump pulse. To this end, we  need to derive  equations of motion for the Bogoliubov quasiparticle densities.
It is advantageous to perform this calculation in the basis in which $H_{\scriptsize\textrm{BCS}}$, Eq.~(\ref{MF_ham}), at $t=t_i$, i.e.,
in the initial state, takes diagonal form. Hence, we consider
the following Bogoliubov transformation
\begin{eqnarray} \label{BdgTrafo}
 a^{\phantom{\dag}}_{\vec{k} l \auf}
 &=&
 u^{ }_{{\bf k} l }\alpha^{\phantom{\dag}}_{\vec{k} l }-v^{\phantom{*}}_{ {\bf k} l }\beta^\dagger_{ \vec{k} l} ,
\; \;
 a^\dagger_{ -\vec{k} l \downarrow }
 =
 v^{\ast}_{{\bf k} l }\alpha^{\phantom{\dag}}_{ \vec{k} l}+u^{\ast}_{{\bf k} l }\beta^\dagger_{\vec{k} l},
\hspace{0.2cm}
\end{eqnarray}
where the Bogoliubov coefficients $u_{{\bf k} l}$ and $v_{{\bf k} l}$ are time-independent, with
$u_{{\bf k} l}=\sqrt{1/2
\left(
1+ \varepsilon_{\vec{k}l} / E_{\vec{k} l}
\right)
}$
and $v_{{\bf k} l}=e^{i\phi_{l}}\sqrt{
1/2
\left(
1-   \varepsilon_{\vec{k}l} / E_{\vec{k} l}
\right)}$,
where $E_{\vec{k}l}=\sqrt{{ \varepsilon^2_{\vec{k}l}}+\mid \Delta_{l} ( t_i ) \mid^2}$.
The phase factors of the superconducting gaps in the initial state $e^{i\phi_{l}} = \Delta_{l} (t_i) / \mid\Delta_{l } (t_i) \mid$
are chosen to be either $(+1)^l$ or $(-1)^l$, depending on whether $V_{12}$ in the initial state is attractive or repulsive.
Substituting Eq.~(\ref{BdgTrafo}) into Eq.~(\ref{MF_ham}) yields
\begin{eqnarray}
H_{\scriptsize\textrm{BCS}} (t)
&=&
\sum_{\vec{k} \in \mathcal{W}  , l}
\Big[
\eta_{1 \vec{k} l} (t)
\left(
\alpha^{\dagger}_{\vec{k} l}\alpha^{\phantom{\dag}}_{\vec{k}l}
+\beta^{\dagger}_{\vec{k}l}\beta^{\phantom{\dag}}_{\vec{k}l }
\right)
\nonumber\\
&& \qquad
+
\eta_{2 \vec{k} l} (t)
\alpha^{\dagger}_{\vec{k} l}\beta^{\dagger}_{\vec{k} l}
+
\eta^{\ast}_{2 \vec{k} l} (t)
\beta^{\phantom{\dag}}_{\vec{k} l} \alpha^{\phantom{\dag} }_{\vec{k} l}
\Big] ,
\end{eqnarray}
where
$\eta^{\phantom{\ast}}_{1\vec{k} l} (t)
=
\frac{1}{E_{\vec{k} l} }
\Big\{
\varepsilon^2_{\vec{k} l}
+ \textrm{Re} \left[
\Delta_{l }^{*}(t) \Delta_{l} ( t_i )
\right]
\Big\} $,
and
$\eta^{\phantom{\ast}}_{2 \vec{k} l} (t)
=
\Delta_{l} ( t_i)
\left\{
\frac{\varepsilon_{\vec{k} l}   }{E_{\vec{k} l}}
\left(
\textrm{Re}
\left[
\frac{\Delta_{l }(t)}{\Delta_{l} ( t_i ) }
\right]
\hspace{-0.1cm} -   \hspace{-0.1cm} 1 \right)
+
i \textrm{Im}
\left[
\frac{\Delta_{l}(t)}{\Delta_{l } (t_i) }
\right]
\right\}$.
Note that here, as before,
all terms except $\Delta_l ( t)$ are time-independent.
In particular, we have assumed that $\varepsilon_{\vec{k} l}$  has
negligible  time-dependence.
Combining Eqs.~(\ref{MFgaps}) and~(\ref{BdgTrafo}) we find
that the superconducting order parameters at time $t$ can be expressed
as
\bea \label{BCS-noneq}
\Delta_{l} (t)
&=&
\hspace{-0.2cm}
\sum_{\vec{k} \in \mathcal{W} , l'}
V_{ll'}
\Big[
v_{\vec{k} l'}^2  \bra \beta^{\dagger}_{\vec{k} l'}\alpha^{\dagger}_{\vec{k} l'}\ket (t)
- u_{\vec{k} l'}^2 \bra  \alpha^{\phantom{\dag}}_{ \vec{k} l' }  \beta^{\phantom{\dag}}_{ \vec{k} l'} \ket (t)
\nonumber\\
&&  \hspace{-0.55cm}  +
u^{\phantom{\ast}}_{\vec{k} l' } v^{\phantom{\ast}}_{\vec{k} l'}
\left( \bra \alpha^{\dagger}_{\vec{k} l'}\alpha^{\phantom{\dag}}_{\vec{k} l' }\ket (t)
+\bra \beta^{\dagger}_{\vec{k} l'}\beta^{\phantom{\dag}}_{\vec{k} l'}\ket (t) -1 \right)
\Big] .
\eea
Hence, the temporal evolution of the  gaps is fully determined
by the time-dependence  of the Bogoliubov quasiparticle densities
$\bra\alpha^{\dagger}_{\vec{k} l}\alpha^{\phantom{\dag}}_{\vec{k}' l}\ket$,
$\bra\beta^{\dagger}_{\vec{k} l}\beta^{\phantom{\dag}}_{\vec{k}' l}\ket$,
$\bra\alpha^{\phantom{\dag}}_{\vec{k}l}\beta^{\phantom{\dag}}_{\vec{k}' l}\ket$,
and
$\bra\beta^{\dagger}_{ \vec{k} l}\alpha^\dagger_{ \vec{k}' l }\ket$.
As shown in the Appendix, it is
straightforward to derive a closed set of
equations of motion
for these quantities, see Eq.~(\ref{EOMeins}).

We numerically solve the equations of motion for the Bogoliubov quasiparticle densities (see Appendix)
to determine the time evolution of the order parameter amplitudes $\left| \Delta_l (t) \right| $.
As is evident from Eq.~(\ref{EOMeins}),
off-diagonal elements in the quasiparticle density matrices, such	as, e.g., $\langle \alpha^{\dag}_{\vec{k} l} \alpha^{\phantom{\dag}}_{\vec{k}+n \vec{q}_0l} \rangle$,
are of order $\left| \vec{A}_0 \right|^n$.
Thus, for sufficiently small $|\vec{A}_0|$, the off-diagonal entries decrease rapidly as $n$ increases.
Therefore we set all off-diagonal elements with $n > 4$ to zero, which substantially reduces the computational costs.
For the numerical simulations we choose the BCS ground state at zero temperature as the initial state.
In what follows, we consider two model systems,
one with weak attractive $V_{12}$, relevant for MgB$_2$, and the other with strong repulsive $V_{12}$,
relevant for iron-based superconductors.

We start by discussing the model appropriate for the multiband superconductor
MgB$_2$. The electronic structure of MgB$_2$ consists of two sets of bands:
$\sigma$-bands forming quasicylindrical hole pockets and $\pi$-bands
giving rise to  tubular networks of electronlike Fermi surface sheets.
Superconductivity in MgB$_2$ arises from couplings between the electrons
and the E$_{2g}$-phonon mode.
These couplings lead to attractive intraband interactions,
which are stronger in the $\sigma$-bands than in the $\pi$-bands.
Since the interband pair scattering is weak, this gives rise to
two gaps with different magnitudes on the  $\sigma$- and $\pi$-bands.

We describe the band structure of MgB$_2$ in a somewhat oversimplified manner
by two parabolic bands\cite{mgb2data}, i.e, a hole band, with Fermi velocity
$v^{\sigma}_{\scriptsize\textrm{F}} = 0.273 \times 10^6$ m/s and effective mass $m_{\sigma} = - 3 m_0$,
and an electron band, with $v^{\pi}_{\scriptsize\textrm{F}} = 1.0 \times 10^6$ m/s and
$m_{\pi} = m_0$. Here, $m_0$ denotes the free electron mass.
The intraband pairing interactions on the $\sigma$- and $\pi$-bands
are chosen to be $V_{11}= 24.8$ meV and $V_{22} = 11.6$ meV, respectively,
whereas the attractive interband coupling is taken to be much smaller than either
$V_{11}$ or $V_{22}$, i.e.,  we set $V_{12}=0.2 V_{11}$.
With these material parameters,
the superconducting gaps in the initial state $\Delta_l ( t_i)$ can
be computed from the self-consistency equation (\ref{gapeq}).
Assuming that the cut-off energy $\hbar \omega_{\scriptsize\textrm{c}}$
is approximately equal to the E$_{2g}$-phonon energy, i.e, $\hbar \omega_{\scriptsize\textrm{c}}=50$ meV,
we find that in the initial state the gaps
on the $\sigma$- and $\pi$-bands are given by  $\Delta_{\sigma} (t_i)  = 7$~meV and  $\Delta_{\pi} (t_i ) = 3$~meV, respectively.
The central energy of the optical pump pulse $\hbar \omega_0$ is chosen to be either in between $2 \Delta_{\pi} (t_i)$ and $2 \Delta_{\sigma} (t_i)$,
or  equal to twice the gap on the $\sigma$-band, i.e., $\hbar \omega_0 = 14$~meV.

Let us first discuss the most interesting case, namely the situation where
the pump photon energy $\hbar \omega_0$ lies in between $2 \Delta_{\pi} (t_i)$ and $2 \Delta_{\sigma} (t_i)$.
In Figs.~\ref{fig1}(a) and \ref{fig1}(b) we plot the calculated time evolution
of $|\Delta_{\sigma} (t) |$ and $|\Delta_{\pi}  (t) |$, respectively, for $\hbar \omega_0 =8$ meV,  $V_{12}=0.2V_{11}$, and three different pump pulse lenghts $\tau=0.4$,
1.0, and $5.0$~ps.
In accordance with previous results on single band superconductors\cite{yuz05,schnyder2011,papenkort,Yuzbashyan06}, we observe two distinct
dynamical regimes: a nonadiabatic regime for pulse widths $\tau$ much smaller than the dynamical time-scales of the superconductor $\tau^{\phantom{*}}_{\Delta_l} \sim h / ( 2 \left | \Delta_l \right| )$  and an adiabatic regime for $\tau \gg \tau^{\phantom{*}}_{\Delta_l}$.
In the nonadiabatic  regime the Bogoliubov quasiparticle densities build up coherently while the system is out of equilibrium. This gives  rise first to a monotonic increase and then to  fast oscillations in the quasiparticle occupations.
Similarly, the superconducting gaps first decrease monotonically and then start to oscillate rapidly (solid green curves in Fig.~\ref{fig1}). Since our model system does not contain any relaxation terms, such as interactions or couplings to an external bath, the oscillations in the quasiparticle occupations are undamped.
 In contrast, the  order parameter oscillations decay as $1/\sqrt{t}$. That is, due to decoherence effects,
the superconducting gap $\Delta_l (t)$ approaches a constant value as time tends to infinity.
To be more precise, we find that to a good approximation
the time evolution of $\left| \Delta_l (t) \right|$ is described by
\begin{eqnarray} \label{twoGapOsc}
|\Delta_{l}(t)|=
 \Delta_{l}^{\infty}
+ \frac{1}{\sqrt{t}}  \sum\limits_{l'=1}^2
b_{ll'} \cos \left( 2 \Delta_{l'}^{\infty}  t+\phi_{ll'} \right) ,
\end{eqnarray}
where $b_{ll'}$ and $\phi_{ll'}$ are constants that depend on the initial conditions.
In other words, the superconducting gaps oscillate with an amplitude decaying as  $1/ \sqrt{t}$
and two frequencies, which are determined by the long-time asymptotic gap values $\Delta^{\infty}_l < \left| \Delta_l ( t_i ) \right|$ [see Figs.~\ref{fig1}(a) and \ref{fig1}(b)].
Interestingly, the relative strength of the two frequency components $\omega_{\Delta_{\sigma}} = 2 \Delta_{\sigma}^{\infty} /\hbar$
 and $\omega_{\Delta_{\pi}} = 2 \Delta_{\pi}^{\infty} /\hbar$ depends sensitively on the magnitude
 of $V_{12}$. That is, the amplitudes $b_{12}$ and $b_{21}$ increase with increasing interband pair scattering $V_{12}$.
 For  $V_{12}=0$, on the other hand, the amplitudes $b_{12}$ and $b_{21}$ vanish, and hence Eq.~(\ref{twoGapOsc}) reduces
 to the well-known result for single band models, where the gap oscillates with a single frequency\cite{yuz05,schnyder2011,papenkort,Yuzbashyan06, pumpPumpProbe}.

\begin{figure}[thp]
  \centering
  \includegraphics[width=0.4\textwidth]{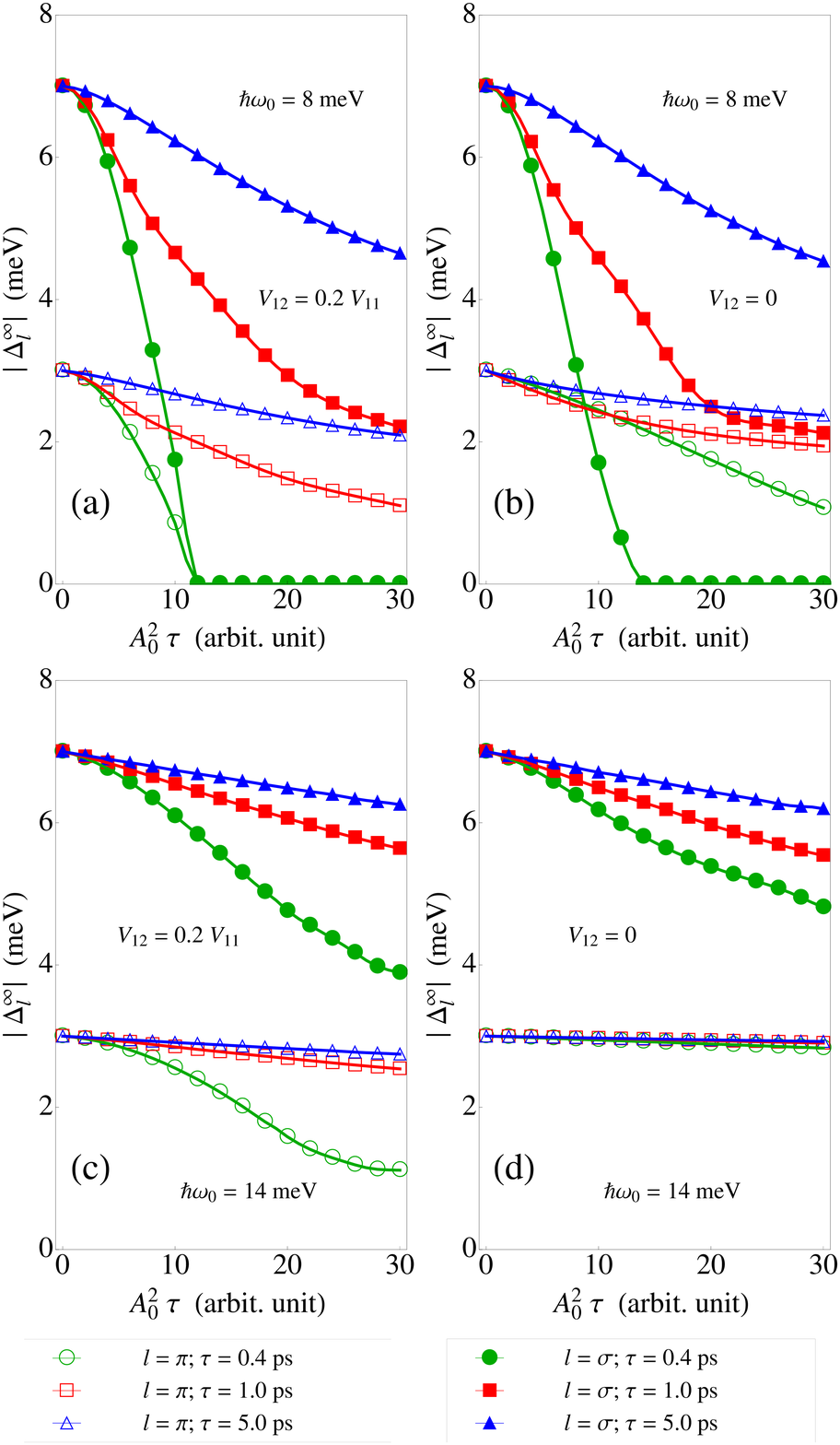}
\caption{ (Color online)
Asymptotic values of the order parameters $\Delta_{\sigma}^{\infty}$  and $\Delta_{\pi}^{\infty}$   as a function of integrated pump pulse intensity for three different pump pulse widths $\tau$.
The central energy of the pump pulse is taken to be
$ \hbar \omega_0 =8$~meV in panels (a) and (b), and $\hbar \omega_0 = 14$~meV in panels (c) and (d).
The left and right panels show the results for $V_{12}=0.2V_{11}$ and $V_{12}=0$, respectively.
The symbols represent  calculated values,
while the solid and dashed lines are a guide to the eye. Here, the arbitrary units in $A_0^2 \tau$  have to be multiplied by the constant $d \approx 1.06 \times 10^{-27} \frac{\textrm{J$^2$s$^3$}}{\textrm{C$^2$m$^2$}}$ to get the physical units.
\label{fig2}
}
\end{figure}

A nonvanishing interband pair scattering $V_{12}$ not only affects the relative strength of the oscillation frequencies but also strongly influences the long-time asymptotic gap values $\Delta_l^{\infty}$ and hence the frequencies of the order parameter oscillations.
This is illustrated in Figs.~\ref{fig2}(a) and \ref{fig2}(b), which show the dependence of $\Delta_l^{\infty}$ on the integrated pump pulse intensity
$A_0^2 \tau$ for $\hbar \omega_0 = 8$~meV and both for zero and nonzero $V_{12}$. Comparing Figs.~\ref{fig2}(a) and \ref{fig2}(b), we observe
that  in the presence of nonzero interband pair scattering  the time evolutions of the two gaps are coupled together.
 The change in $\Delta^{\infty}_{l}$ due to interband interactions seems to be
roughly proportional to  V$_{12} \Delta^{\infty}_{l'}$, with $l' \neq l$ [cf. Eq.~(\ref{BCS-noneq})]. That is, the larger gap $\Delta_{\sigma} (t)$ influences the smaller gap $\Delta_{\pi} (t) $ more strongly than vice versa. With increasing integrated laser intensity both gaps are suppressed simultaneously, until for sufficiently large $A_0^2 \tau$ they  vanish at the same value of $A_0^2 \tau$. For zero $V_{12}$, in contrast, the order parameters can be suppressed independently. For example, we find that in the nonadiabatic regime the larger gap $\Delta_{\sigma}(t)$
vanishes at a smaller integrated intensity than the smaller gap  $\Delta_{\pi}(t) $ [solid green curve in Fig.~\ref{fig2}(b)].
We observe that in general the asymptotic gap values $\Delta_l^{\infty}$ decrease  linearly at small $A_0^2 \tau$, but deviate  from this behavior at larger integrated intensities. With increasing $A_0^2 \tau$, the curves corresponding to short pump pulses ($\tau=0.4$ ps) show a downward bend,
while those with longer pump pulses ($\tau = 1.0$ and $5.0$~ps) flatten due to Pauli blocking.

Let us also briefly discuss the case where the pump pulse energy $\hbar \omega_0$ is equal to $2 \Delta_{\sigma} (t_i) = 14$~meV.
In general, the behavior of the gaps in this case is rather similar to the previously discussed case, where $\hbar \omega_0 = 8$~meV.
As before, we find that  for pump pulse widths $\tau \gg \tau^{\phantom{*}}_{\Delta_l}$ both gap amplitudes $\left| \Delta_l (t) \right|$ decrease
 monotonically from their initial equilibrium values to their final values $\Delta^{\infty}_l$, whereas for $\tau \ll \tau^{\phantom{*}}_{\Delta_l}$ the gaps approach their asymptotic values in an oscillatory fashion according to Eq.~(\ref{twoGapOsc}). However, as can be seen from Figs.~\ref{fig2}(c) and \ref{fig2}(d),
 the dependence of $\Delta^{\infty}_l$ on
 the integrated intensity $A_0^2 \tau$ is quite different than in Figs.~\ref{fig2}(a) and \ref{fig2}(b).
Interestingly, we find that a pump pulse with central energy $\hbar \omega_0 = 2 \Delta_{\sigma} (t_i)$ depletes the superconducting condensate more efficiently on the $\sigma$-band than on the $\pi$-band.  In particular, for $V_{12}=0$ the gap on the $\pi$-band is almost unaffected by the pump pulse, even for large $A_0^2 \tau$, whereas
the asymptotic gap value on the $\sigma$-band decreases steadily with increasing $A_0^2 \tau$.
For nonzero  $V_{12}$ the behavior is similar, although here the gap on the $\pi$-band is slightly suppressed, which is due to the coupling with the gap on the $\sigma$-band.

\begin{figure}
\centering
\includegraphics[width=0.22\textwidth]{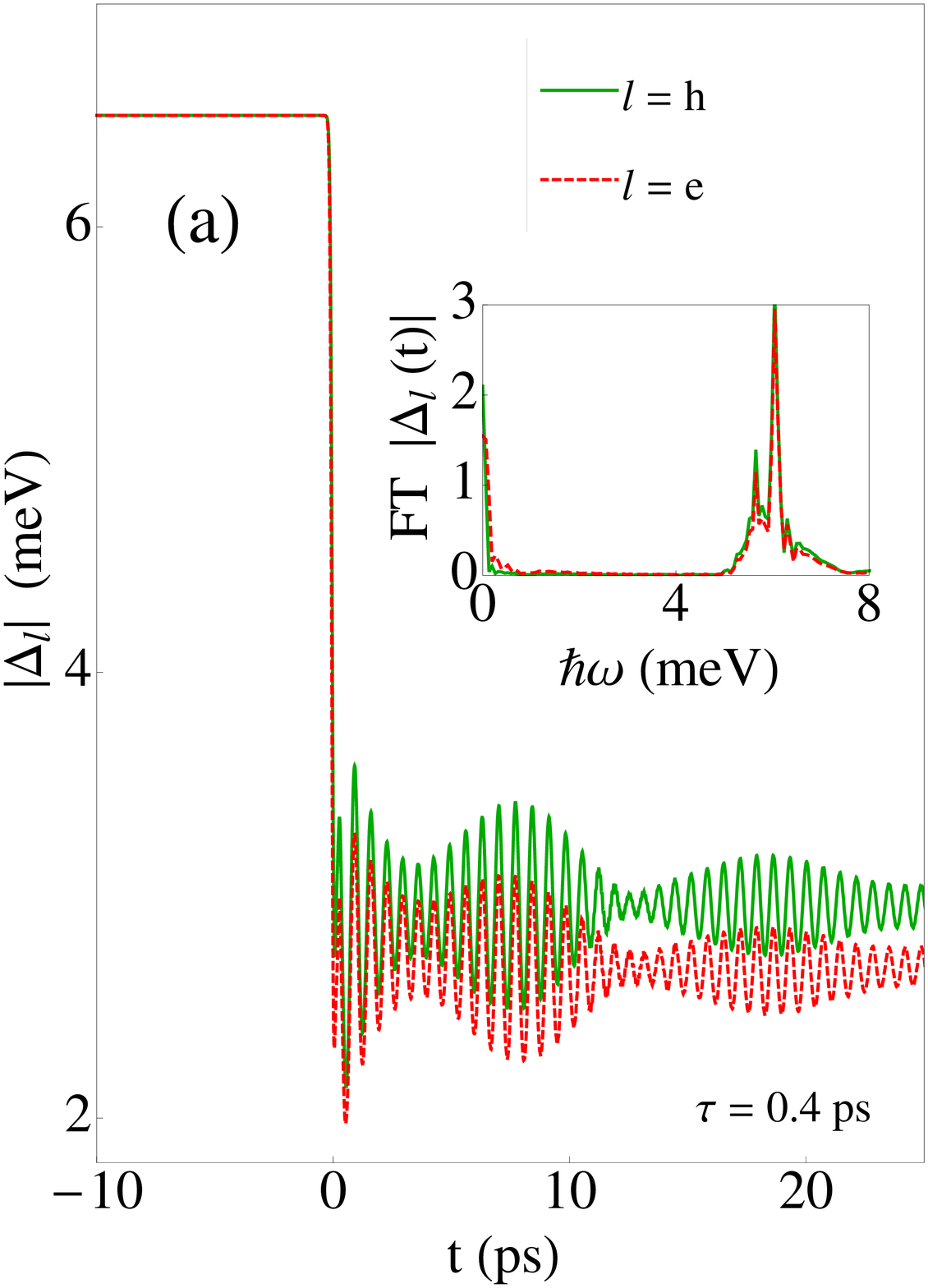}
\includegraphics[width=0.22\textwidth]{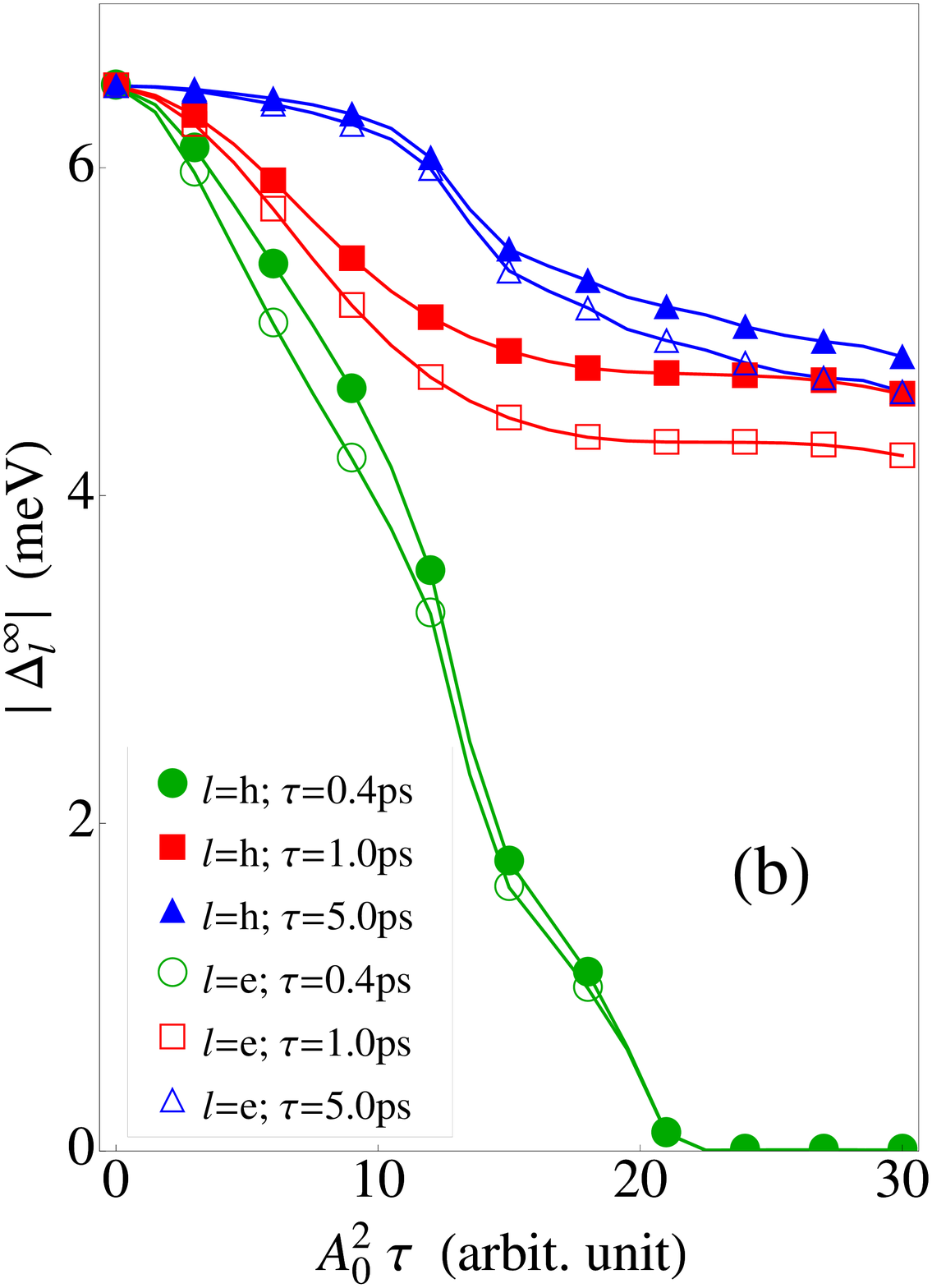}
\caption{(Color online) (a) Calculated time evolution of
$|\Delta_{\scriptsize\textrm{h}}(t) |$  and $|\Delta_{\scriptsize\textrm{e}} (t) |$   for the case of iron-based superconductors.
The inset shows the spectral distribution of the gap oscillations for the same parameters as in the main panel.
(b)~Asymptotic values of the order parameters
$\Delta_{\scriptsize\textrm{h}}^{\infty}$ and $\Delta_{\scriptsize\textrm{e}}^{\infty}$ versus
integrated pump pulse intensity for  three
different pump pulse widths $\tau$.
In this figure we choose $V_{12}=- 10 V_{11}$ and
$\hbar \omega_0 = 14$ meV.
In panel (a), we set $\tau=0.4$ ps and $A^2_0 \tau \approx 1.44 \times 10^{-26} $Js$/$Cm  [corresponding to the point $A^2_0\tau=13$ arb.\ units in panel (b)]. The symbols in panel (b) represent  calculated values, while the lines
are a guide to the eye. Here, the arbitrary units in $A_0^2 \tau$  have to be multiplied by the constant $d \approx 1.11 \times 10^{-27} \frac{\textrm{J$^2$s$^3$}}{\textrm{C$^2$m$^2$}}$ to get the physical units.
}
\label{fig4}
\end{figure}

Secondly, we consider a model appropriate for the iron pnictide superconductors\cite{IronReview}.
Many features of iron-based superconductors can be captured within a simple two-band model
with quasinested holelike and electronlike pockets centered at the $\Gamma$ and $M$ points in the Brillouin zone, respectively \cite{chubukov08}. Due to the quasinesting of the Fermi surfaces,  the repulsive interband pair scattering in such a description is much larger than the intraband pairing potentials. The latter are assumed to be overscreened due to the nesting effects and become attractive for low energies relevant for superconductivity.
Therefore, in what follows, we model the band structure of iron-based superconductors by two parabolic bands, a hole band with effective mass
$m_{\scriptsize\textrm{h}} = -3 m_0$ and Fermi velocity  $v_{\scriptsize\textrm{F}}^{\scriptsize\textrm{h}}=0.168 \times 10^6$ m/s, and an electron band with $m_{\scriptsize\textrm{e}}=m_0$ and $v_{\scriptsize\textrm{F}}^{\scriptsize\textrm{e}}=1.453 \times 10^6$ m/s\cite{pnicdata}.
For convenience we shift the location of the electron Fermi surface from the $M$ point to the $\Gamma$ point, i.e., we
set $\varepsilon_{2{\bf k+Q}}=\varepsilon_{2{\bf k}}$, where ${\bf Q} = (\pi,\pi)$.
The attractive intraband pairing interactions on the electron and hole Fermi surfaces are chosen to be  $V_{11} =  2.466$~meV and $V_{22} = 2.463$ meV,
while the repulsive interband pair scattering is set to
$V_{12}=-24.42$ meV.
With these material parameters and assuming that the  cut-off energy $\hbar \omega_{\scriptsize\textrm{c}}$ is determined by the energy of the spin fluctuations in the paramagnetic state, $\hbar \omega_{\scriptsize\textrm{c}} = 30$~meV\cite{inosov10},
we find that  the equilibrium gaps
are given by $\Delta_{\scriptsize\textrm{h}} (t_i) =-\Delta_{\scriptsize\textrm{e}} (t_i) = 6.5$~meV.
That is, the gaps on the hole and electron bands
are equal in magnitude but opposite in phase, which is roughly in line with experimental findings in some of the hole-doped iron-based superconductors \cite{pnicdata}.
The two-band superconductor is driven out of equilibrium by a pump pulse with energy $\hbar \omega_0 = 14$ meV,
which is of the same order but slightly larger than twice the gap amplitudes in the initial state.

Based on this model for pnictide superconductors, we compute the temporal evolution of the superconducting order parameters on the hole and electron  bands.
In Fig.~\ref{fig4}(a), the time dependence of $\left| \Delta_l (t) \right|$ is shown
for a pump pulse with length $\tau = 0.4$~ps, i.e., for the nonadiabatic regime.
We observe that both gaps exhibit almost the same behavior: they first decrease monotonically and then
start to oscillate with different frequencies producing a beatinglike pattern in the amplitudes.
Due to the smaller Fermi velocity and the larger effective mass of the hole band,
the gap on the hole band $\Delta_{\scriptsize\textrm{h}} ( t)$ is slightly less suppressed than the gap on the electron band $\Delta_{\scriptsize\textrm{e}} (t)$.
Because of the nonzero interband coupling $V_{12}$, both gaps oscillate with two frequencies, $\omega_{\Delta_{\scriptsize\textrm{e}}} = 2 \Delta^{\infty}_{\scriptsize\textrm{e}} / \hbar$
and $\omega_{\Delta_{\scriptsize\textrm{h}}} = 2 \Delta^{\infty}_{\scriptsize\textrm{h}} / \hbar$, which differ very little, i.e., $\left| \omega_{\Delta_{\scriptsize\textrm{h}}}  - \omega_{\Delta_{\scriptsize\textrm{e}}} \right| \ll \omega_{\Delta_{\scriptsize\textrm{e}}}$.
This in turn leads to a pronounced beating phenomenon, as can be seen in Fig.~\ref{fig4}(a).

The dependence of the asymptotic gap values $\Delta^{\infty}_{l}$ on the integrated pump pulse intensity $A^2_0 \tau$
is plotted in Fig.~\ref{fig4}(b). Due to the large interband coupling $V_{12}$ the gaps on both bands decrease
almost synchronously with increasing  $A^2_0 \tau$. However, the asymptotic value of the gap on the electron band $\Delta^{\infty}_{\scriptsize\textrm{e}}$  is always slightly
smaller than the asymptotic value of the gap on the hole band $\Delta^{\infty}_{\scriptsize\textrm{h}}$. As discussed above, this is because of the larger effective mass
and the smaller Fermi velocity on the hole band. In general, we find that the relative difference between $\Delta^{\infty}_{\scriptsize\textrm{h}}$ and $\Delta^{\infty}_{\scriptsize\textrm{e}}$
gradually increases with increasing pump pulse intensity.
Correspondingly, the difference between the two oscillation frequencies $ \omega_{\Delta_{\scriptsize\textrm{h}}} $ and $ \omega_{\Delta_{\scriptsize\textrm{h}}} $  increases,
and hence the beating phenomenon becomes less pronounced, as
the integrated pump pulse intensity is increased.


In this work we analyzed the nonequilibrium dynamics of two-band superconductors after excitation by a short opitcal pump pulse.
We considered two model systems: one with dominant intraband pairing and weak attractive interband pair scattering,
and one where the repuslive intrerband interactions are much larger than the intraband pairing potentials.
The former model is relevant for MgB$_2$ superconductors, whereas the latter one is appropriate for iron-based superconductors.
For both of these model systems we numerically computed the time evolution of the order parameters
as a function of pump pulse duration and integrated pump pulse intensity. Our main observation is that the ratio between the gaps for asymptotically large times depends sensitively on the interband Cooper-pairing strength and differs from its equilibrium value. This allows in principle to use pump-probe experiments for direct identification of the
interband Cooper-pair scattering strength. 
In addition, sufficiently short pump pulses create fast oscillations in the gap amplitudes and the quasiparticle densities of these two-band superconductors. We found that for nonzero interband pair scattering $V_{12}$
these oscillations are characterized by two frequencies which are determined by the long-time asymptotic
values of the gaps on the two bands (see Fig.~\ref{fig1}).
We showed that the relative strength of the two frequency components sensitively depends on the magnitude of the interband interaction $V_{12}$. When the gaps on the two bands are of similar magnitude
(which is the case, for example, in iron-based superconductors) the relative
difference between the two oscillation frequencies is small, and hence the
gaps oscillate with a beatinglike pattern in the amplitudes.

\subsection{Acknowledgments}
We thank A.~Avella, U.~Bovensiepen, B.~Kamble, N.~Hasselmann, and G.~Uhrig  for useful discussions. The work of A.A. and I.E. is supported by the Merkur Foundation.


%
\subsection{Appendix}
In this Appendix, we present  one of the equations of motion (as an example)  which determine the time evolution of the quasiparticle densities
$\bra\alpha^{\dagger}_{\vec{k} l}\alpha^{\phantom{\dag}}_{\vec{k}' l}\ket$,
$\bra\beta^{\dagger}_{\vec{k} l}\beta^{\phantom{\dag}}_{\vec{k}' l}\ket$,
$\bra\alpha^{\phantom{\dag}}_{\vec{k}l}\beta^{\phantom{\dag}}_{\vec{k}' l}\ket$,
and
$\bra\beta^{\dagger}_{ \vec{k} l}\alpha^\dagger_{ \vec{k}' l }\ket$.
Here, as in the main text, $l=1,2$ denotes the band index.
Since $\bra\alpha^{\phantom{\dag}}_{\vec{k}l}\beta^{\phantom{\dag}}_{\vec{k}' l}\ket$
and
$\bra\beta^{\dagger}_{ \vec{k}' l}\alpha^\dagger_{ \vec{k} l }\ket$
are related by Hermitian conjugation, only six out of the eight quasiparticle densities need to be evaluated numerically.
By use of Heisenberg's equation of motion,  the following equation for $\bra\alpha^{\dagger}_{\vec{k} l}\alpha^{\phantom{\dag}}_{\vec{k}' l}\ket$
can be obtained
\bea \label{EOMeins}
&&
i\hbar\frac{ d}{dt}\bra\alpha^{\dagger}_{ \vec{k} l}\alpha^{\phantom{\dagger}}_{\vec{k}' l}\ket
=
(\eta_{1 \vec{k}' l}-\eta_{1 \vec{k} l})
\bra \alpha^{\dagger}_{ \vec{k} l}
\alpha^{\phantom{\dagger}}_{\vec{k}' l }
\ket
\nonumber
\\
\nonumber
&&-
\eta_{2 \vec{k}' l}
\bra
\beta^{\dagger}_{\vec{k}' l}
\alpha^{\dagger}_{ \vec{k} l}
\ket
+
\eta^*_{2  \vec{k} l}
\bra
\alpha_{ \vec{k}' l}
\beta_{ \vec{k}  l}
\ket
      -\frac{e\hbar}{m_l}
\sum_{\vec{q} =\pm \vec{q}_0}
           \vec{k}\cdot\vec{A}_{\vec{q} }
\\
\nonumber
&&{\Big (}
M^{+}_{l \vec{k}' ,\vec{k}'-\vec{q} }
\bra
\alpha^{\dagger}_{\vec{k} l }\alpha_{ \vec{k}'-\vec{q}  l}
\ket
 +
 L^{-}_{l \vec{k}',\vec{k}'-\vec{q} }
 \bra
 \beta^{\dagger}_{ \vec{k}'-\vec{q} l}\alpha^{\dagger}_{ \vec{k} l}
 \ket
 \\
\nonumber
&&          -M^{+}_{l \vec{k} +\vec{q} ,\vec{k} }
           \bra
           \alpha^{\dagger}_{ \vec{k} +\vec{q} l }\alpha_{\vec{k}' l}
           \ket
           -L^{-*}_{l \vec{k} ,\vec{k} +\vec{q} }
           \bra
           \alpha_{\vec{k}' l}\beta_{ \vec{k} +\vec{q} l}
           \ket
{\Big )}
\\\nonumber
&& \qquad
+  \frac{e^2}{2m_l}
\sum_{\vec{q}} \left( \sum_{\vec{q}\prime = \pm \vec{q}_0}
\vec{A}_{\vec{q} -\vec{q}\prime }\cdot\vec{A}_{\vec{q}\prime }
 \right)
\\
\nonumber
&&
\Big(
M^{-}_{l \vec{k}',\vec{k}'-\vec{q} }
\bra
\alpha^{\dagger}_{ \vec{k} l}\alpha_{ \vec{k}'-\vec{q} l }
\ket
 +L^{+}_{l \vec{k}'-\vec{q} ,\vec{k}'}
 \bra
 \beta^{\dagger}_{ \vec{k}'-\vec{q} l }\alpha^{\dagger}_{ \vec{k} l}
 \ket
\nonumber\\
&&
           -M^{-}_{l \vec{k} +\vec{q} ,\vec{k} }
           \bra
           \alpha^{\dagger}_{ \vec{k} +\vec{q} l }\alpha_{\vec{k}' l}
           \ket
           -L^{+*}_{l \vec{k} ,\vec{k} +\vec{q} }
           \bra
           \alpha_{\vec{k}' l}\beta_{ \vec{k} +\vec{q} l}
           \ket
\Big),
\label{EOF1}
\nonumber\\
&&
\eea
where
we have introduced the short-hand notation
$M^{\pm}_{l \vec{k},\vec{k}^\prime}=u^{\phantom{\ast}}_{\vec{k} l}u^{*}_{\vec{k}^\prime l}
\pm v^{\phantom{\ast}}_{\vec{k}l}v^{*}_{\vec{k}^\prime l}$ and  $L^{\pm}_{l\vec{k},\vec{k}^\prime}= u_{\vec{k} l}v_{\vec{k}^\prime l}
\pm v_{\vec{k} l}u_{\vec{k}^\prime l}$,
with the coherence factors $u_{\vec{k} l}$ and $v_{\vec{k} l}$.
The functions $\eta_{1 \vec{k} l}$ and $\eta_{2  \vec{k} l}$
are defined in the main text.
In deriving Eqs.~(\ref{EOF1}), 
we used the relation  $\vec{A}^*_{\vec{q}}=\vec{A}^{\phantom{*}}_{-\vec{q}}$.


\end{document}